\documentclass[prd,showpacs,floatfix,twocolumn,superscriptaddress
]{revtex4}

\usepackage[dvips]{graphicx}
\usepackage{amssymb,amsmath}
\usepackage{times}

\newcommand{\be}{\begin{equation}}
\newcommand{\ee}{\end{equation}\noindent}
\newcommand{\bea}{\begin{eqnarray}}
\newcommand{\eea}{\end{eqnarray}}

\newcommand{\maprightb}[1]{\smash{\mathop{
\hbox to 1cm{\rightarrowfill}}\limits_{#1}}}

\newcommand{\bc}{\begin{center}}
\newcommand{\ec}{\end{center}}

\newcommand{\matTwo}{\left(\begin{array}{rr}}
\newcommand{\matThree}{\left(\begin{array}{rrr}}
\newcommand{\emat}{\end{array}\right )}
\newcommand{\detTwo}{\left|\begin{array}{rr}}
\newcommand{\detThree}{\left|\begin{array}{rrr}}
\newcommand{\edet}{\end{array}\right |}

\newcommand{\bra}{\langle}
\newcommand{\ket}{\rangle}
\newcommand{\tr}{{\rm tr}}

\def\thline{\noalign{\hrule height 1pt}}

\begin{document}
\title{Imaginary Chemical Potential Approach for the Pseudo-Critical Line in 
the QCD Phase Diagram  with Clover-Improved Wilson Fermions}

\author{Keitaro Nagata}
\email[]{nagata@rcnp.osaka-u.ac.jp}
\affiliation{
{\textit Department of Physics, The University of Tokyo,
Bunkyo-ku, Tokyo 113-0033 JAPAN }
}
\affiliation{
{\textit Research Institute for Information Science and Education, Hiroshima University, Higashi-Hiroshima 739-8527 JAPAN}
}
\author{Atsushi Nakamura}
\email[]{nakamura@riise.hiroshima-u.ac.jp}
\affiliation{
{\textit Research Institute for Information Science and Education, Hiroshima University, Higashi-Hiroshima 739-8527 JAPAN}
}

\date{\today}

\begin{abstract}
The QCD phase diagram is studied in the lattice QCD simulation 
with the imaginary chemical potential approach. 
We employ a clover-improved Wilson fermion action of two-flavors and a 
renormalization-group improved gauge action, and perform the simulation 
at an intermediate quark mass on a $8^3\times 4$ lattice. 
The QCD phase diagram in the imaginary chemical potential $\mu_I$ region is 
investigated by performing the simulation for more than 150 points on 
the $(\beta,\mu_I)$ plane. 
We find that the Roberge-Weiss phase transition at $\mu_I/T=\pi/3$ is 
first order and its endpoint is second order, which are identified by 
the phase of the Polyakov loop. We determine the pseudo-critical line 
from the susceptibility of the Polyakov loop modulus. 
We find a clear deviation from a linear dependence of the pseudo-critical 
line on $\mu_I^2$. 
\end{abstract}


\maketitle

\begin{figure*}[htbp]
\begin{center}
\includegraphics[width=0.35\linewidth]{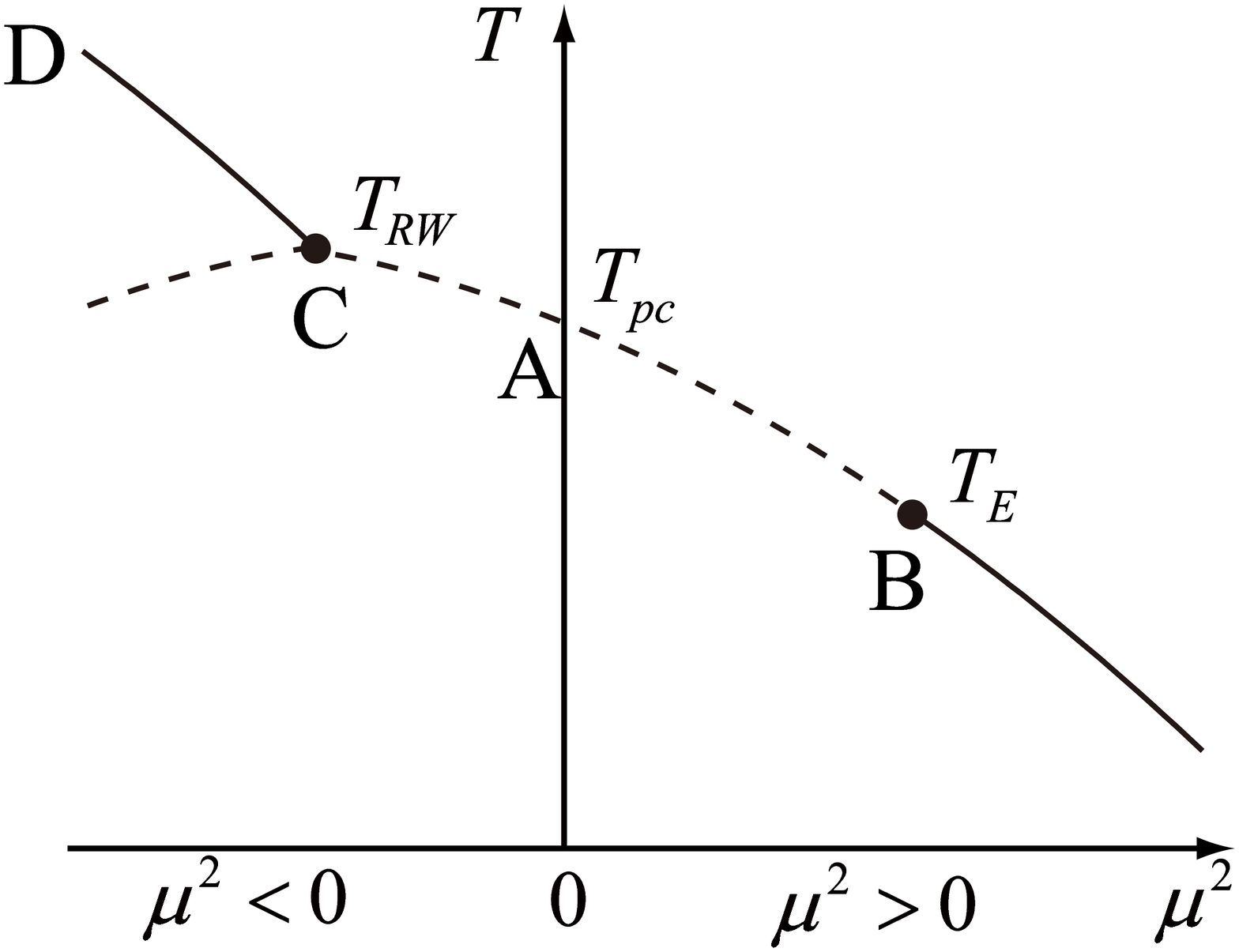}
\includegraphics[width=0.35\linewidth]{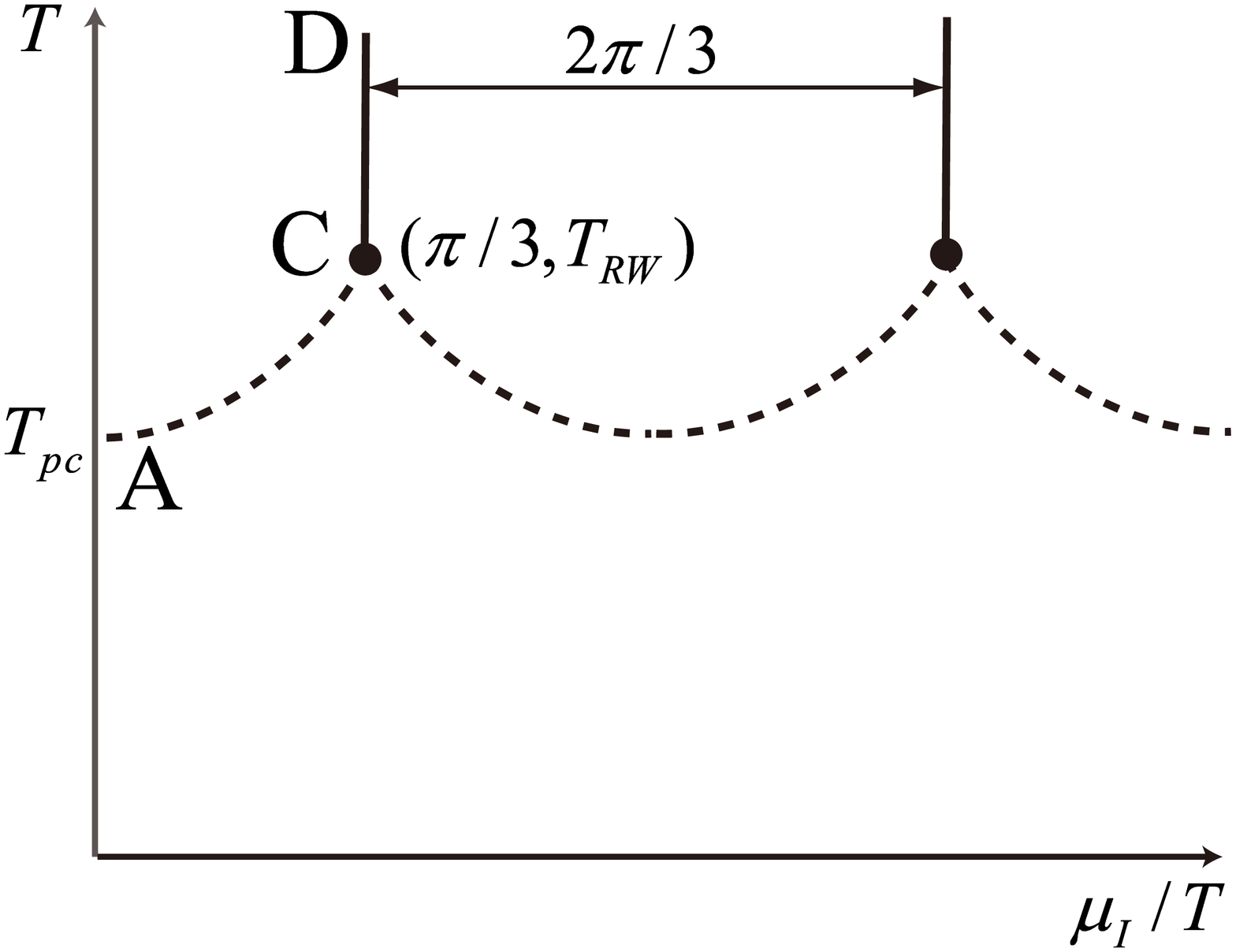}
\begin{minipage}{0.75\linewidth}
\caption{Schematic figures for the $N_f=2$ QCD phase diagram in 
the $(\mu^2, T)$ plane (left) and $(\mu_I/T, T)$ plane (right). 
A : Pseudo-critical point at $\mu=0$. B : Critical endpoint. 
C : Roberge-Weiss endpoint. 
AB : Pseudo-critical line. AC : Extension of the line AB into the 
imaginary chemical potential plane. CD : Roberge-Weiss phase 
transition line $\mu_I/T=\pi/3$. In the right panel, larger 
$\mu_I/T$ region of the phase diagram is obtained from  the RW periodicity.
}\label{Jan2311fig1}
\end{minipage}
\end{center}
\end{figure*}
\section{Introduction}

The QCD phase diagram, which includes states of matter formed in 
terms of the strong interaction, has been of prime interest in recent 
physics covering particle physics, hadron/nuclear physics and astrophysics. 
Because QCD is non-perturbative in most regions of the QCD phase diagram, 
one is forced to use the lattice QCD in order to obtain a quantitative 
understanding. The lattice QCD is expected to provide reliable information on 
the phase structure based on QCD. Indeed, recently, there have been many active 
quantitative investigations about the finite temperature QCD~\cite{Fodor:2010,DeTar:2011nm}. 

On the other hand, simulations of systems with non-zero quark chemical 
potential $\mu$ have been a long challenge for the lattice QCD because 
of the notorious sign problem. 
In the lattice QCD, a fermionic determinant $\det \Delta(\mu)$ is used as 
a probability in a Monte Carlo method. The introduction of non-zero $\mu$ 
makes $\det \Delta(\mu)$ complex, and therefore leads to the breakdown of the 
stochastic part of the lattice QCD, see Ref.~\cite{Muroya:2003qs}.

Despite of the severe sign problem, several approaches have been 
proposed to study the QCD with nonzero $\mu$, see 
e.g.~\cite{Muroya:2003qs,deForcrand:2010ys}. One idea is to avoid 
the sign problem by performing simulations in systems with an imaginary 
chemical potential. A partition function and its free-energy are analytic 
within one phase even if chemical potential is extended to complex, which 
is true until the occurance of a phase transition.
This validates the imaginary chemical potential approach for the study of the 
QCD phase diagram.

Fermion determinants satisfy a well-known relation 
\begin{align}
\Delta(\mu)^\dagger = \gamma_5 \Delta( - \mu^*) \gamma_5, 
\label{Jan1111eq1}
\end{align}
which holds for a complex chemical potential:
 $\mu= \mu_R + i\mu_I (\mu_R, \mu_I\in\mathbb{R})$. 
 This implies that $\det \Delta(\mu)$ is complex for a real chemical potential 
$\mu=\mu_R$, which causes the sign problem. 
On the other hand, one can easily prove $\det \Delta(\mu)$ is real 
for a pure imaginary chemical potential $\mu= i\mu_I$.
The sign problem does not occur in this case, and Monte Carlo methods 
are available. The imaginary chemical potential approach provides an 
insight into the QCD phase diagram through the analytic continuation. 
In addition, data obtained in such a simulation can be used for the matching 
of phenomenological models such as Polyakov loop extended 
Nambu-Jona-Lasinio(PNJL) models with the lattice QCD~\cite{Sakai:2008py,Kouno:2009bm}. 

The imaginary chemical potential approach has been studied by using 
staggered fermions with two flavor~\cite{deForcrand:2002ci,D'Elia:2009qz,D'Elia:2009tm}, 
three flavor~\cite{deForcrand:2010he}, four 
flavor~\cite{D'Elia:2002gd,D'Elia:2004at,D'Elia:2007ke,Cea:2010md} 
in 2-color QCD and finite isospin QCD~\cite{Cea:2009ba,Cea:2007vt}, by using
Wilson fermions with two flavor~\cite{Wu:2006su}. 

Staggered fermions of the standard type might have suffered from two problems. 
First, it needs a fourth-root trick for one flavor~\footnote{
At zero and imaginary chemical potential a square-root is enough for studying $N_f=2$.}.
Second, it does not show a scaling behavior expected from three-dimensional 
O(4) spin models for the finite temperature transition~\cite{AliKhan:2000iz}, 
although the possibility of the first oder phase transition for 
two degenerate flavors was studied in Ref.~\cite{Bonati:2009yg}.
Wilson fermions are free from the fourth root tricks and show the correct 
scaling behavior. On the other hand, Wilson fermions suffer from an 
explicit breaking of chiral symmetry. However, one can define a subtracted 
chiral condensate, which satisfy a correct scaling behavior. 
Although Wilson fermions require more computational time 
than that required in staggered fermions, simulations with Wilson fermions 
are now possible even on the physical quark masses at zero density.

In finite temperature simulations with the combination of the plaquette gauge
action and the standard Wilson quark action at $N_t=4$, 
the transition is smooth crossover at small and large quark mass and 
rapid crossover at intermediate quark mass~\cite{Bernard:1993en}, which 
is different from what is expected the transition becomes sharp for light 
and heavy quark masses. This unexpected behavior is removed by 
improving the gauge action~\cite{Iwasaki:1996ya,AliKhan:2000iz}. 
The improvement of the gauge action is essential in removing lattice artifacts
at finite lattice spacings. 

Thus, the study of the Wilson fermions with improved terms is complementary and
useful to confirm results obtained in other actions~\cite{deForcrand:2002ci,D'Elia:2009qz,D'Elia:2009tm,deForcrand:2010he,D'Elia:2002gd,D'Elia:2004at,D'Elia:2007ke,Wu:2006su} 
and to establish a better understanding of the QCD phase diagram.  
In this paper, we study the two-flavor QCD phase diagram at an intermediate 
quark mass by using the imaginary chemical potential approach.  
This is the first employment of the two-flavor Wilson fermion 
with a clover term and the renormalization-group(RG) improved 
gauge action to the imaginary chemical potential approach.
 
This paper is organized as follows. In the next section, we briefly review 
properties and issues of the imaginary chemical potential region of the 
phase diagram. The setup for the simulation is also explained here. 
The numerical results are shown in Sec.~\ref{Feb2711sec1}. 
We investigate the deconfinement transition and Roberge-Weiss endpoint  
in Sec.~\ref{Mar0111sec1}, and Roberge-Weiss phase transition line 
in Sec.~\ref{Mar0111sec2}. We determine the pseudo-critical line in 
Sec.~\ref{Mar0111sec3}. The final section is devoted to a summary. 

\section{Framework}

\subsection{Phase diagram with imaginary chemical potential} 
We begin with a brief overview of the QCD phase diagram and 
of issues in question. 

First we show in the left panel of Fig.~\ref{Jan2311fig1} an expected 
phase diagram in $(\mu^2, T)$ plane containing both the real 
$(\mu^2\ge 0, \mu=\mu_R)$ and imaginary $(\mu^2\le 0, \mu=i\mu_I)$ regions. 
Even if $\mu^2\le 0$, it is expected quark-gluon-plasma(QGP) 
and hadronic phases exist at high and low  temperatures, respectively. 
The two phases are separated by the deconfinement crossover transition line, 
which is an extension from the $\mu^2 \ge 0$ region. 
These are consequences of an analyticity of the grand partition function. 
The absolute value of the Polyakov loop is often employed to identify 
confinement/deconfinement phase, although it is not a real order
parameter because of the crossover nature of the transition.

Two characteristics of the $\mu^2\le 0$ region are so-called Roberge-Weiss(RW) 
phase transition and Roberge-Weiss(RW) periodicity~\cite{Roberge:1986mm}. 
The QCD grand partition function has a periodicity with a period $2\pi/N_c$ as 
\begin{align} 
Z\left(\frac{\mu_I}{T}\right)= Z\left(\frac{\mu_I}{T} + \frac{2\pi k }{N_c}\right), 
\end{align} 
where  $k$ is an integer. Furthermore, Roberge and Weiss showed from 
a perturbative analysis the existence of a first-order phase transition 
on the line $\mu_I/T=\pi/N_c$, and from a strong coupling analysis 
the absence of such a transition at low temperatures. These features hold for 
SU($N_c$) gauge theories. Hereafter we consider $N_c=3$. 
The RW phase transition relates to the $Z(3)$ 
symmetry and an order parameter identifying this phase transition is the phase or 
imaginary part of the Polyakov loop. 
Because the RW phase transition occurs at high temperatures but does not at low temperatures, 
it may have an endpoint at a temperature $T_{RW}$ on the line $\mu_I/T=\pi/3$. 
These features are well manifested in the $(\mu_I/T, T)$-phase diagram, see the 
right panel of Fig.~\ref{Jan2311fig1}. 

\begin{figure}[htbp] 
\begin{center} 
\includegraphics[width=0.9\linewidth]{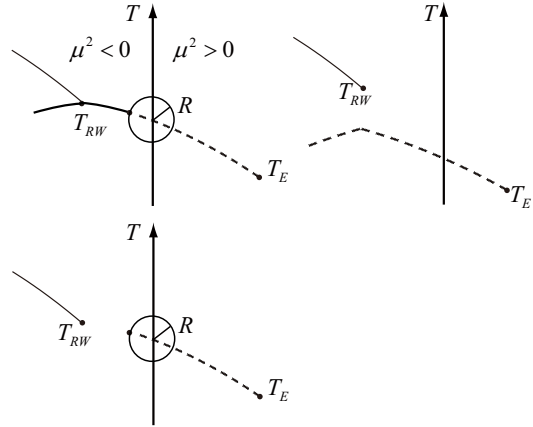} 
\begin{minipage}{0.75\linewidth} 
\caption{Counter examples against the naive expectation. 
$R$ is a convergence radius of the pseudo-critical line determined 
in the $\mu^2\le 0$ region.
}\label{Mar0311fig1} 
\end{minipage} 
\end{center} 
\end{figure} 

Although the phase diagrams in Fig.~\ref{Jan2311fig1} are naively expected, 
several points remain as issues which should be discussed further. 
Figure \ref{Jan2311fig1} is drawn according to the two points :
\begin{itemize}
\item The RW phase transition exists at high temperatures and has an endpoint.
\item The extension of the crossover line exists in the $\mu^2\le 0$ region.
\end{itemize}
The second point needs an assumption that a pseudo-critical line can be 
defined for a crossover. Once the pseudo-critical line is defined, the second point 
is ensured by the identity theorem. 

Based on the above two points, one can consider several counter examples, 
against the naive expectation Fig.~\ref{Jan2311fig1}, see  Fig.~\ref{Mar0311fig1}. 
In order to reveal the phase structure without assumptions on imagination, 
we need to know
\begin{itemize}
\item the location and order of the RW endpoint
\item the location of phase boundary of the deconfinement transition
\item the way with which the RW and deconfinement transition lines are connected 
\end{itemize}
Indeed, the quark-mass dependence of the RW endpoint was found in 
Ref.~\cite{D'Elia:2009qz,deForcrand:2010he}. D'Elia et al discussed ~\cite{D'Elia:2009qz}
a possibility that other first-order phase transition lines depart from the RW endpoint, 
which corresponds to the left top panel in Fig~\ref{Mar0311fig1}.

The above questions are important for two reasons. 
First, they are relevant with the definition range of the pseudo-critical line 
determined by the imaginary chemical potential approach, which also relates to the 
applicable range of the line in the $\mu^2\ge 0$ region.
For instance, if the pseudo-critical line has an endpoint near $T_{pc}$ at $\mu=0$~\footnote{
Here we do not mean that a pseudo-critical line has an  endpoint defined in 
statistical mechanics. Rather, we use a term  endpoint for a situation where
a change between the confinement and deconfinement phases becomes very continuous, 
and the pseudo-critical temperature is not well-defined.}, 
then a convergence radius $R$ of the pseudo-critical line 
obtained from the imaginary chemical potential approach is given by a distance between 
the endpoint and $T_{pc}$. The pseudo-critical line obtained can be applied to 
a domain with the convergence radius. 

Second, it is speculated from the Lee-Yang theorem~\cite{Yang:1952be,Lee:1952ig}
that a phase transition in the $\mu^2\le 0$ region is relevant
to one in the $\mu^2\ge 0$ region, although the distribution of 
the Lee-Yang zeros of the QCD grand partition function has not 
been well understood. Assuming the Lee-Yang zeros of the QCD 
are distributed on a line in the complex fugacity plane, it is 
possible that the RW phase transition line and its endpoint in 
the $\mu^2\le 0$ region reflect the first order phase transition 
line and critical endpoint in the $\mu^2\ge 0$ region of the QCD 
phase diagram. 

\subsection{Formulation and setup}
\begin{figure*}[htbp]
\begin{center}
\includegraphics[width=0.45\linewidth]{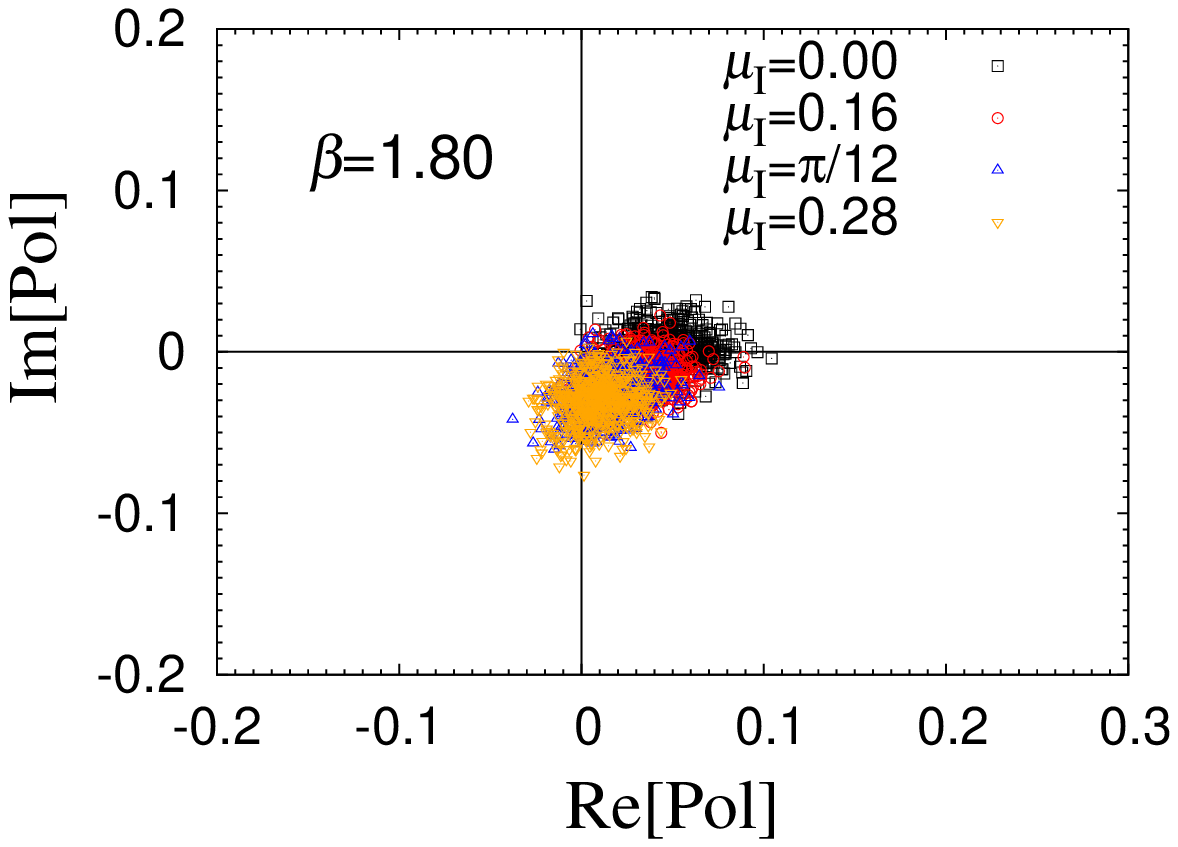}
\includegraphics[width=0.45\linewidth]{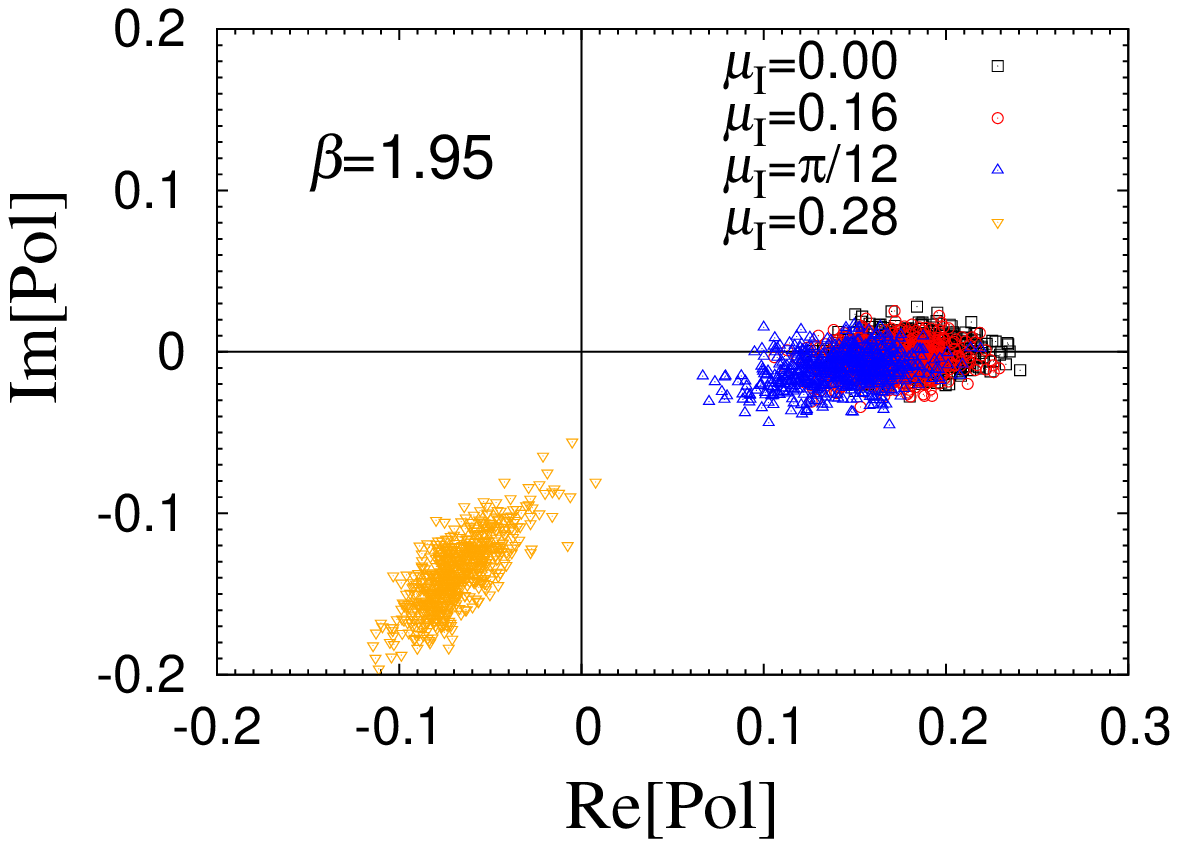}
\begin{minipage}{0.75\linewidth}
\caption{Scatter plots of the Polyakov loop. Left : $\beta=1.80$ 
(low temperature (below $T_{pc}$)).  Right : $\beta=1.95$ (high temperature (above $T_{RW}$)). 
}\label{Mar0411fig1}
\end{minipage}
\end{center}
\end{figure*}
\begin{figure*}[htbp]
\begin{center}
\includegraphics[width=0.45\linewidth]{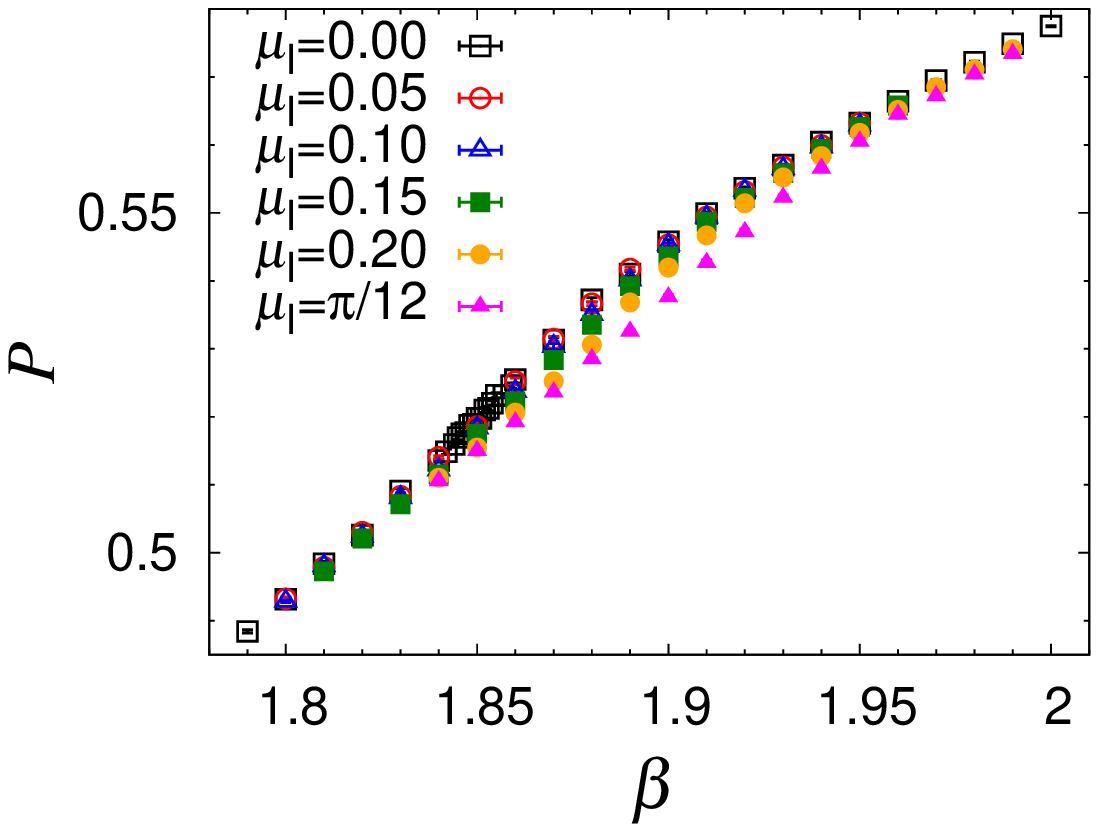}
\includegraphics[width=0.45\linewidth]{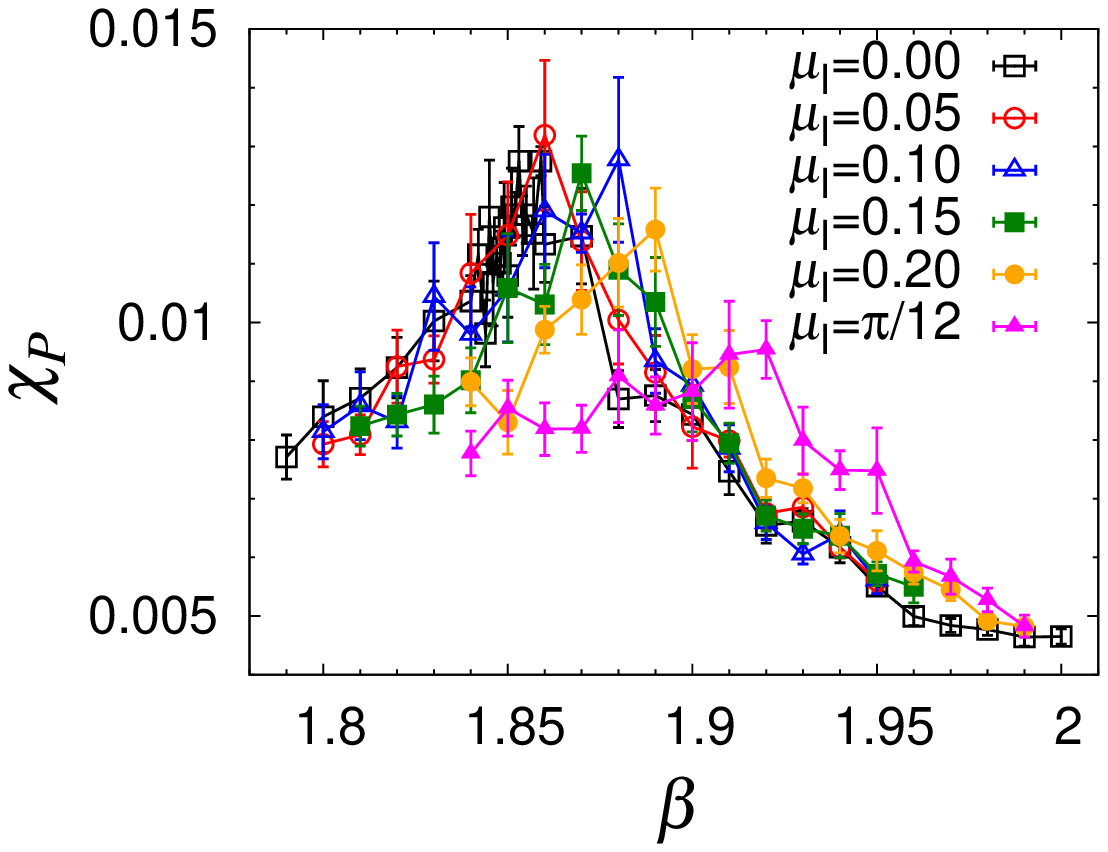}
\begin{minipage}{0.8\linewidth}
\caption{The $\beta$-dependence of the plaquette $P$ (left) 
and its susceptibility $\chi_P$ (right) for various $\mu_I$. 
}\label{Jan1011fig1}
\end{minipage}
\end{center}
\end{figure*}
\begin{figure*}[htbp]
\begin{center}
\includegraphics[width=0.45\linewidth]{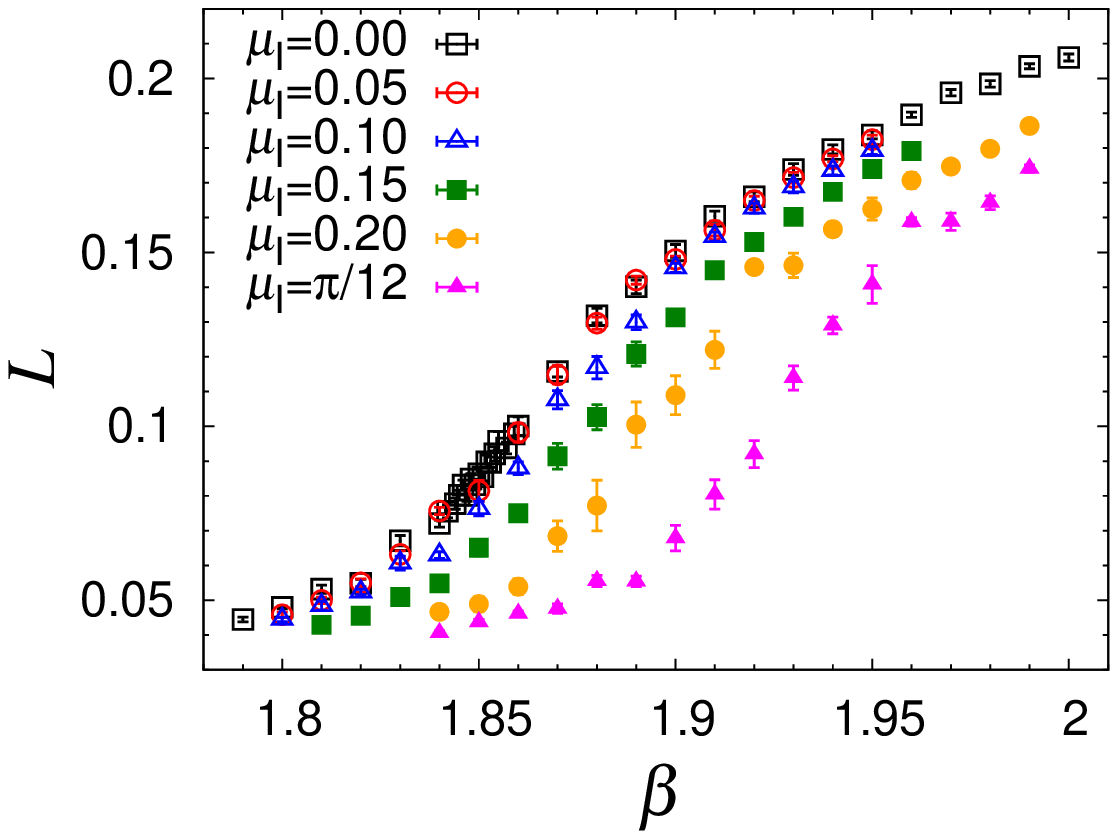}
\includegraphics[width=0.45\linewidth]{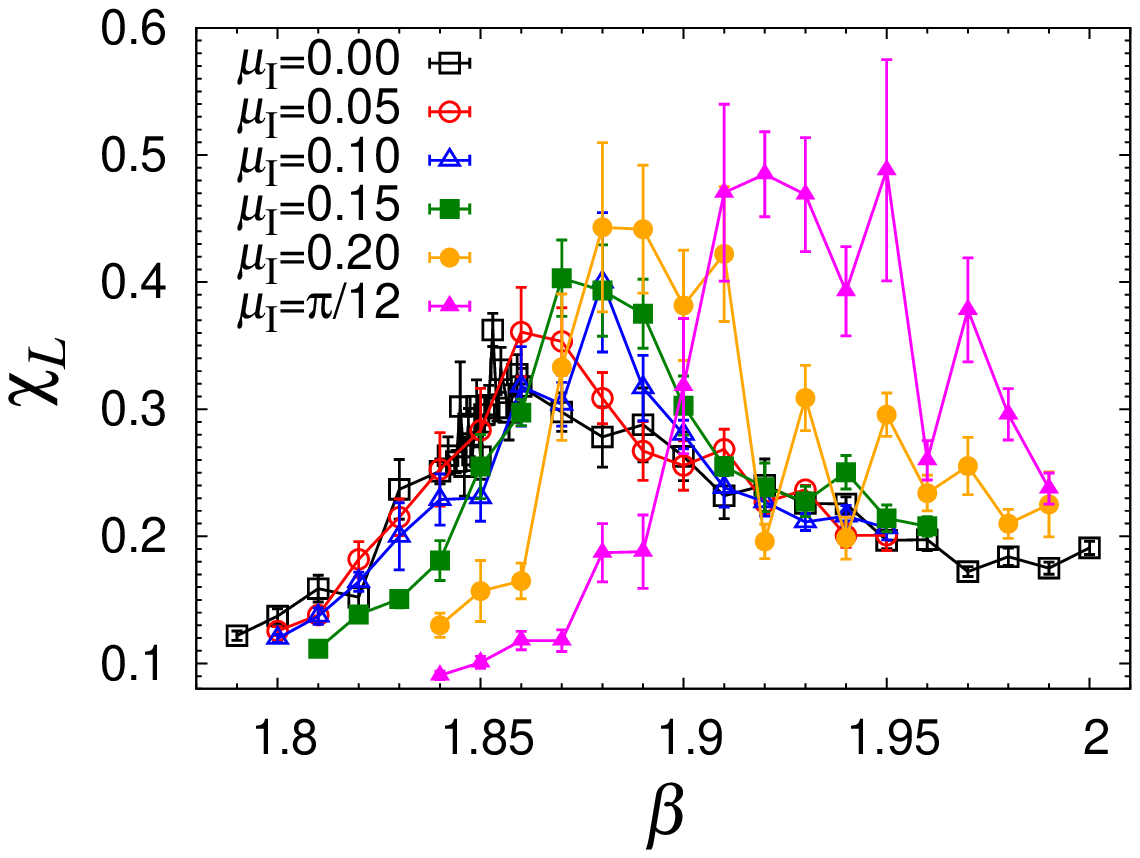}
\begin{minipage}{0.75\linewidth}
\caption{The $\beta$-dependence of the Polyakov loop modulus $L$ (left) and 
its susceptibility $\chi_L$ (right) for various $\mu_I$. 
}\label{Jan1011fig2}
\end{minipage}
\end{center}
\end{figure*}
\begin{figure*}[htbp]
\begin{center}
\includegraphics[width=0.45\linewidth]{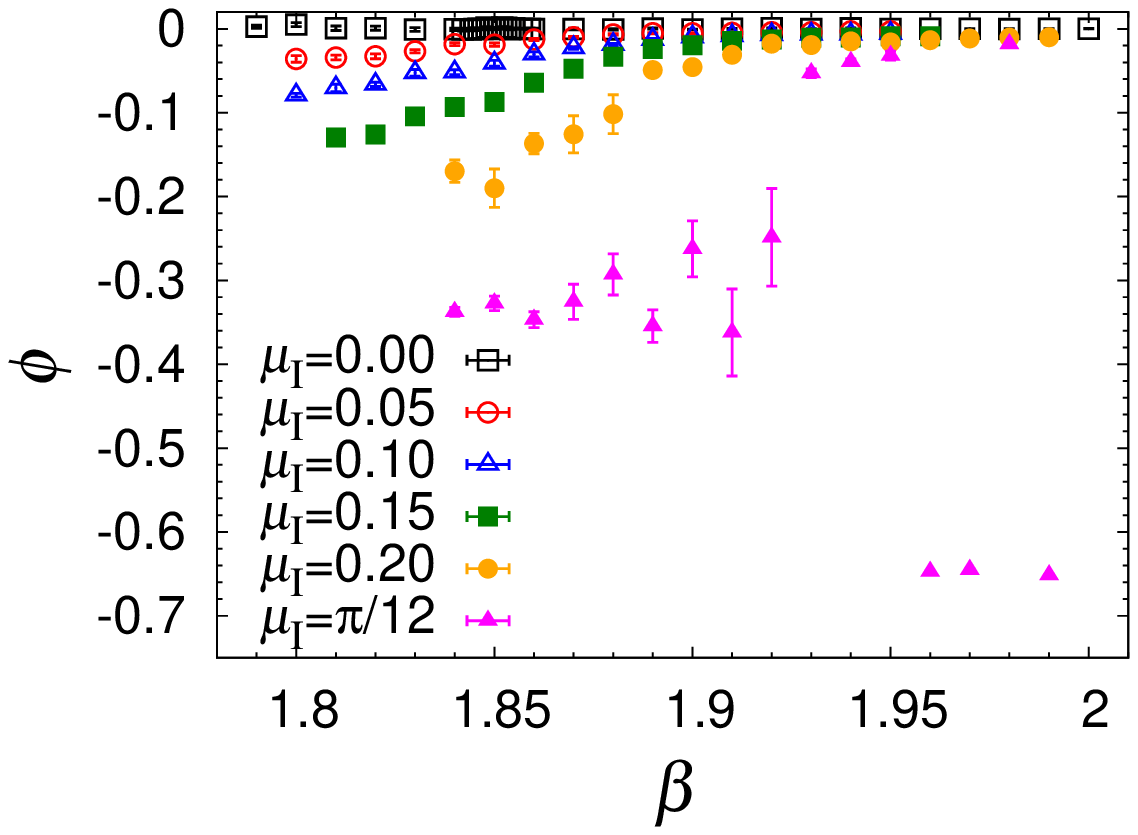}
\includegraphics[width=0.45\linewidth]{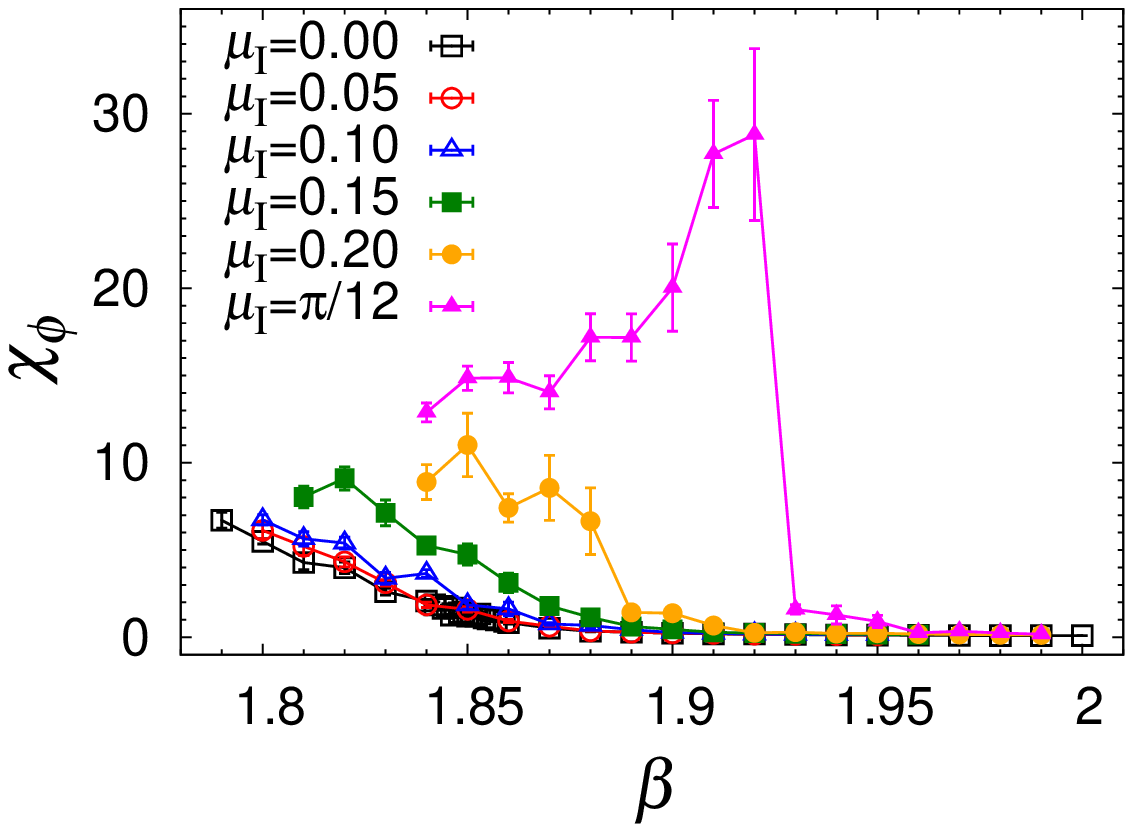}
\begin{minipage}{0.75\linewidth}
\caption{The $\beta$-dependence of the Polyakov loop phase $\phi$ (left) and 
its susceptibility $\chi_\phi$ (right) for various $\mu_I$.
}\label{Jan1011fig3}
\end{minipage}
\end{center}
\end{figure*}

We employ the RG-improved gauge action~\cite{Iwasaki:1985we}
\begin{eqnarray}
S_g = \frac{\beta}{6} \left[ c_0 \sum_{x,\mu<\nu} W_{\mu\nu}^{1\times 1}(x)  
+ c_1 \sum_{x,\mu, \nu} W_{\mu\nu}^{1\times 2}(x) \right], 
\end{eqnarray}
with $c_1 = -0.331$ and $c_0 = 1-8 c_1$, and 
the clover-improved Wilson fermion action with the quark matrix 
\begin{widetext}
\begin{eqnarray}
\Delta(x,y) =  \delta_{x, x^\prime} 
&-&\kappa \sum_{i=1}^{3} \left[
(1-\gamma_i) U_i(x) \delta_{x^\prime, x+\hat{i}} 
+ (1+\gamma_i) U_i^\dagger(x^\prime) \delta_{x^\prime, x-\hat{i}}\right] \nonumber \\
 &-&\kappa \left[ e^{+\mu} (1-\gamma_4) U_4(x) \delta_{x^\prime, x+\hat{4}}
+e^{-\mu} (1+\gamma_4) U^\dagger_4(x^\prime) \delta_{x^\prime, x-\hat{4}}\right] \nonumber \\
&-& \kappa  C_{SW} \delta_{x, x^\prime}  \sum_{\mu \le \nu} \sigma_{\mu\nu} 
F_{\mu\nu}.
\end{eqnarray}
\end{widetext}
Here $\mu$ is the quark chemical potential in lattice unit, which is introduced to 
the temporal part of link variables.

In order to scan the phase diagram, simulations were done 
for more than 150 points on the $(\mu_I, \beta)$ plane in the domain 
$0\le \mu_I \le  0.28800$ and $1.79 \le \beta \le 2.0$. 
Note that the RW phase transition line in the present setup is given by 
$\mu_I = \pi/12\sim 0.2618$. 
All the simulations were performed on a $N_s^3\times N_t = 8^3\times 4$ lattice. 
The value of the hopping parameter $\kappa$ were determined for each value of $\beta$ 
according to a line of the constant physics with $m_{PS}/m_V=0.8$ obtained in 
Ref.~\cite{Ejiri:2009hq}. The coefficient of the clover term $C_{SW}$ was
determined by using a result obtained in the one-loop perturbation 
theory~\cite{Sheikholeslami:1985ij} : $C_{SW} = ( 1- 0.8412 \beta^{-1})^{-3/4}$. 

The hybrid Monte Carlo algorithm were employed to generate gauge configurations. 
The setup for the molecular dynamics was as follows: a step size $\delta \tau = 0.02$, 
number of the molecular dynamics $N_{\tau}=50$ and length  $N_{\tau}\delta\tau =1$. 
The acceptance ratio for this setup was more than 90 \%. We generated 11, 000 
trajectories for most parameter sets, and 16, 000 trajectories for some parameter 
sets near the deconfinement transition at $\mu_I=0$. For all the ensemble, the 
first 5,000 trajectories were removed as thermalization. The plaquette $P$, 
Polyakov loop $L e^{i\phi}$ and their susceptibilities were measured for each 
trajectory, where $L$ and $\phi$ are the modulus and phase of the Polyakov loop. 

The Polyakov loop operator is as usual defined by 
\begin{align}
P_{ol} = \frac{1}{N_V N_c} \sum_{x} \tr \prod_{t=1}^{N_t}U_4(\vec{x}, t)_,
\end{align}
where $N_V=N_s^3$. The modulus and phase are, after the ensemble average, defined by 
$\bra P_{ol}\ket = L e^{i\phi}$. Their susceptibilities are also defined by 
\begin{align}
\chi_L &= N_V \bra (L - \bra L \ket )\ket^2 , \\
\chi_\phi &= N_V \bra (\phi - \bra \phi \ket )\ket^2 . 
\end{align}

Distributions of the Polyakov loop in the complex plane are 
shown in Fig.~\ref{Mar0411fig1}. As the figures clearly show, 
the phase structure can be identified  by considering the $\beta$ and 
$\mu_I$ dependence of the Polyakov loop. 

\section{Numerical Results}
\label{Feb2711sec1}

\subsection{Deconfinement transition and RW endpoint}
\label{Mar0111sec1}

First, we investigate the deconfinement transition and RW 
endpoint, by considering the $\beta$-dependence of the observables.

The plaquette $P$ and its susceptibility $\chi_P$ are shown in 
Fig.~\ref{Jan1011fig1}. $P$ is a smooth increasing function of 
$\beta$ for all $\mu_I$. The effect of $\mu_I$ suppresses $P$ 
for intermediate $\beta$, while does not change it for small and large $\beta$. 
However, the effect is up to a few percent. 
$\chi_P$ has a broad peak, and the peak position moves toward 
larger $\beta$ with increasing $\mu_I$. 

The Polyakov loop modulus $L$ and its susceptibility $\chi_L$ 
are shown in Fig.~\ref{Jan1011fig2}. $L$ increases slowly for small 
$\beta$. At a certain value of $\beta$, the slope of $L$ becomes large, 
and $\chi_L$ has a broad peak. 
These behaviors suggest the possibility that the system undergoes the 
crossover transition with increasing $\beta$ or temperature. 
The peak position tends to move towards a larger $\beta$ 
with increasing $\mu_I$, similar to the behavior of $\chi_P$. 
This behavior confirms that pseudo-critical temperatures become higher with
the increase of $\mu_I$ until $\mu_I=\pi/12$. 

It should be noted that the crossover behavior is observed for all 
$\mu_I$, even on the line $\mu_I=\pi/12$. Hence, the pseudo-critical 
line starts from $\mu_I=0$ and reaches $\mu_I=\pi/12$.

Also note that the peak of $\chi_P$ and $\chi_L$ are not sharp, and 
the signal is unclear. This may come from the small spatial and temporal sizes. 
In fact, the WHOT collaboration reported in a finite temperature 
simulation with the same action that the hopping parameter dependence of the Polyakov loop 
susceptibility shows a pronounced peak in a $16^3 \times 6$ lattice~\cite{Ejiri:2009hq}.  

\begin{figure}[htbp]
\begin{center}
\includegraphics[width=\linewidth]{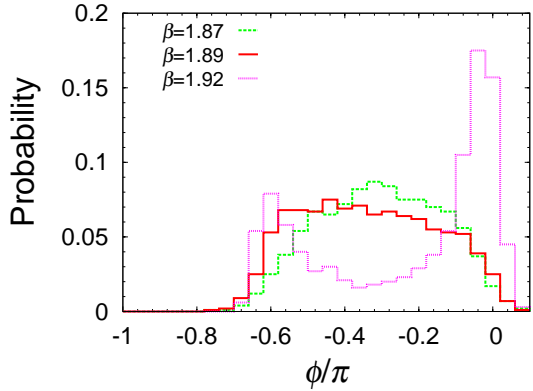}
\begin{minipage}{0.75\linewidth}
\caption{The histogram of $\phi$ at $\mu_I=\pi/12$ for various $\beta$.
}\label{Jan2111fig1}
\end{minipage}
\end{center}
\end{figure}
The Polyakov loop phase $\phi$ and its susceptibility $\chi_\phi$ 
are shown in Fig.~\ref{Jan1011fig3}. $\phi$ rapidly changes near 
$\beta=1.92$ only for $\mu_I=\pi/12$, while it is a smooth function 
of $\beta$ for $0\le \mu_I < \pi/12$. It is seen  that for $\mu_I=\pi/12$ 
there is one vacuum at low temperatures and are two vacua at high temperatures. 
The histogram of $\phi$ at $\mu_I=\pi/12$ in Fig.~\ref{Jan2111fig1} also 
shows this behavior. 
The susceptibility $\chi_\phi$ shows a divergent-like behavior near 
$\beta=1.92$ only for $\mu_I=\pi/12$. 
These behaviors suggest the possibility that the system undergoes the second order 
phase transition at the RW endpoint with increasing temperature.  

The transition point of $\chi_\phi$ in Fig.~\ref{Jan1011fig3} agrees with 
the the peak position of $\chi_L$ in Fig.~\ref{Jan1011fig2} within error bars.
Hence, the pseudo-critical line is connected with the Roberge-Weiss phase 
transition line at the RW endpoint. This is the case shown in Fig.~\ref{Jan2311fig1}. 
We observed that near the RW endpoint, the Polyakov loop modulus shows 
the crossover-like behavior and its phase shows the second order-like behavior. 
Note that the order of the phase transition were naively obtained from the behaviors 
of the observables. 
In order to confirm the order of the phase transition, the finite volume 
scaling analysis should be investigated in future works.

\subsection{RW phase transition}
\label{Mar0111sec2}

\begin{figure}[htbp]
\begin{center}
\includegraphics[width=0.95\linewidth]{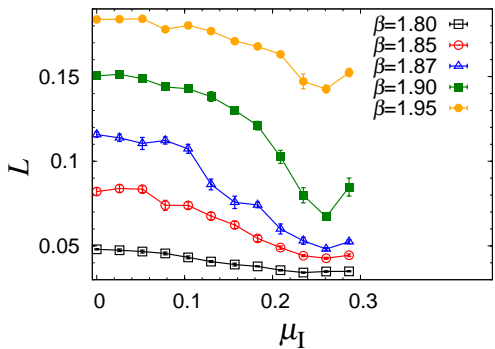}
\begin{minipage}{0.8\linewidth}
\caption{The $\mu_I$-dependence of $L$ for various $\beta$.
}\label{Jan2111fig4}
\end{minipage}
\end{center}
\end{figure}

\begin{figure*}[htbp]
\begin{center}
\includegraphics[width=0.45\linewidth]{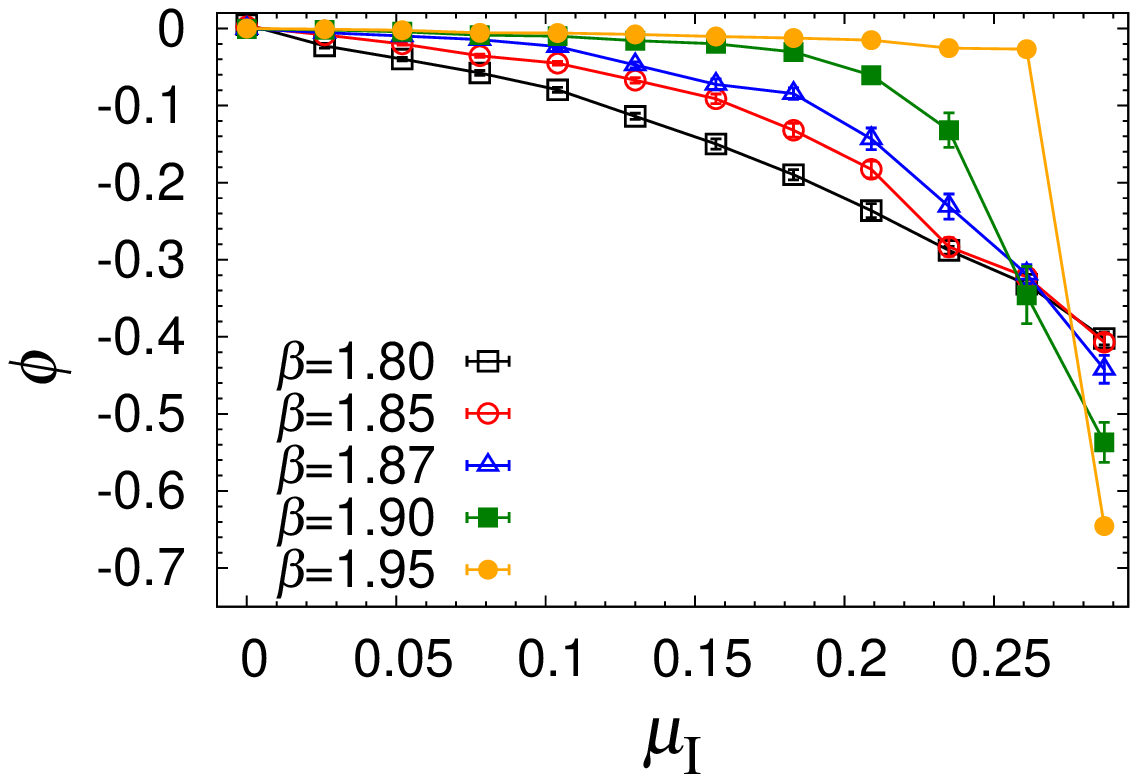}
\includegraphics[width=0.45\linewidth]{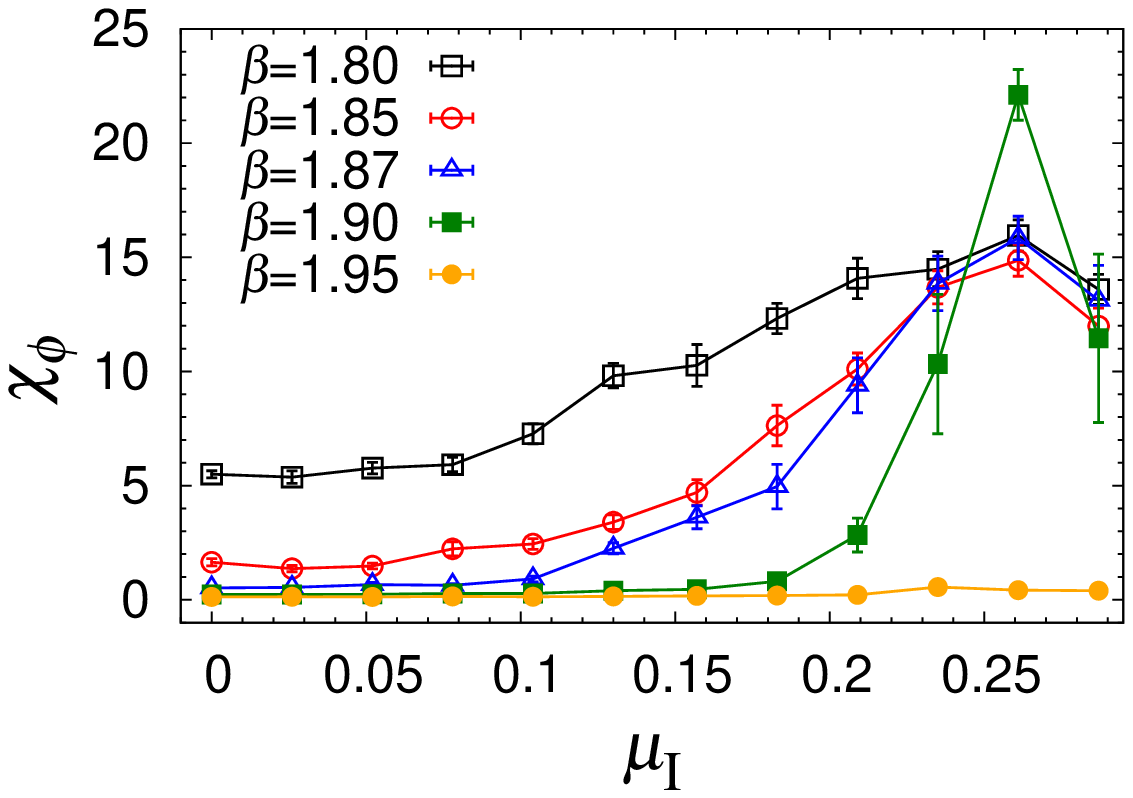}
\begin{minipage}{0.75\linewidth}
\caption{The $\mu_I$-dependence of $\phi$ and $\chi_\phi$ for various $\beta$.
}\label{Jan2111fig5}
\end{minipage}
\end{center}
\end{figure*}

Next, we show the $\mu_I$-dependence of the observables in 
Figs.~\ref{Jan2111fig4} and \ref{Jan2111fig5}, and investigate the nature of 
the RW phase transition line. Results of $P$ are not shown there, because 
the effect of $\mu_I$ changes the value of $P$ up to a few percent, 
as shown in the previous subsection. 

$L$ decreases for $\mu_I<\pi/12$, and increases for $\mu_I>\pi/12$. 
A rapidly decreasing behavior, which occurs due to the intersection 
with the pseudo-critical line, is observed for $\beta=1.87$ and $1.90$. 
The line for $\beta=1.95$ crosses the RW phase transition line at 
$\mu_I=\pi/12$. However, a critical behavior is not observed there. 

The $\mu_I$-dependence of $\phi$ and $\chi_\phi$ are shown in Fig.~\ref{Jan2111fig5}. 
$\phi$ is a smooth function of $\mu_I$ at low temperatures ($\beta=1.80$-$1.90$), 
while $\phi$ jumps to $-2\pi/3$ from $0$ at $\mu_I=\pi/12$ at a high temperature 
($\beta=1.95)$. The system undergoes the first order phase transition at $\mu_I=\pi/12$ 
at high temperatures. Note $L$ is periodic and $\phi$ is anti-periodic, which is caused by 
the periodicity of the $\mu_I$-dependence of the Polyakov loop~\cite{Kouno:2009bm}.

Together with the result obtained in the previous subsection, we find that 
the RW phase transition is the first order one and ends at second order 
endpoint, which are identified by the phase of the Polyakov loop. 
This feature is consistent with the results obtained in Ref.~\cite{deForcrand:2002ci,Wu:2006su}. 
It was reported~\cite{D'Elia:2009qz,deForcrand:2010he} that the order of the RW endpoint 
depends on the quark mass, and first order for light and heavy quark masses and second order 
for intermediate quark masses. Hence the second order nature of the RW endpoint comes from 
the intermediated quark mass. 

\subsection{Pseudo-critical line}
\label{Mar0111sec3}
As we have discussed in Sec.~\ref{Mar0111sec1}, the Polyakov loop modulus 
$L$ shows the deconfinement crossover with increasing temperature, 
which spans the range $0\le \mu_I \le \pi/12$. 
We extract the value of $\beta_{pc}$ by fitting five or six data of $\chi_L$ near 
the peak with a Gaussian function: $\chi_L \propto \exp(-b(\beta-\beta_{pc})^2)$. 
\begin{table}[htbp]
\begin{center}
\begin{minipage}{0.75\linewidth}
\caption{The values of $\beta_{pc}$ determined by fitting $\chi_L$ with 
a Gaussian function. 
} \label{Jan2111tab1}
\vspace{0.5cm}
\end{minipage}
\begin{tabular}{llcl}
\thline
$\mu_I$ & $\beta_{pc}$ & $\delta \beta_{pc}$\\
\hline 
0.00 & 1.866  & 0.007   \\
0.05 & 1.866  & 0.001   \\
0.10 & 1.877  & 0.008   \\
0.15 & 1.880  & 0.002   \\
0.20 & 1.891  & 0.001   \\
0.26 & 1.927  & 0.005   \\
\thline
\end{tabular}
\end{center}
\end{table}

\begin{figure}[htbp]
\begin{center}
\includegraphics[width=\linewidth]{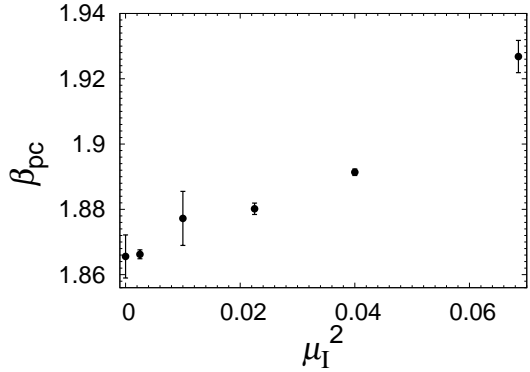}
\begin{minipage}{0.75\linewidth}
\caption{$\beta_c$ and $\mu_I^2$. 
}\label{Jan2511fig1}
\end{minipage}
\end{center}
\end{figure}
The results are shown in Table~\ref{Jan2111tab1}, and 
plotted as a function of $\mu_I^2$ in Fig.~\ref{Jan2511fig1}.
The data points in Fig.~\ref{Jan2511fig1} clearly deviate from a linear function 
near $\mu_I\sim \pi/12$, which implies the contributions of higher-order terms. 
The need of terms of order higher than $\mu^2$ in the
functional form of the pseudo-critical line was first pointed out in 
\cite{Cea:2007vt,Cea:2009ba,Cea:2010md}. 
Such a behavior has not been obtained from a study with the combination of 
plaquette gauge action and the standard Wilson fermion without the clover term~\cite{Wu:2006su}. 
Hence the clover-improved Wilson fermion and the RG-improved gauge action 
leads to the deviation from a linear dependence of the pseudo-critical 
line on $\mu_I^2$. This finding is an advantage of the improved actions. 

In general, the pseudo-critical line can be expanded in powers of $\mu_I$ : 
\begin{align}
\beta_{pc} (\mu_I) = \sum_{n} c_n (\mu_I^2)^n,
\label{Mar0611eq1}
\end{align} 
which is defined for the range $0\le \mu_I \le \pi/12$ bounded by the RW endpoint. 
Within the definition range and with the present numerical results, only a few terms 
can be determined. Here, we test quadratic and quartic functions
and a Pad\'e approximation of a simple type
\begin{align}
\beta_{pc}(\mu_I) =  c_0 \frac{ 1 + c_1 \mu_I^2}{1 + c_2 \mu_I^2}.
\label{Mar0611eq2}
\end{align}

\begin{table}[htbp]
\begin{center}
\begin{minipage}{\linewidth}
\caption{The coefficients of the fit functions for the pseudo-critical line. 
$c_0$ is fixed with the central value of $\beta_{pc}(0)$. 
}\label{Feb2811tab1}
\vspace{0.5cm}
\end{minipage}
\begin{tabular}{lccccc}
\thline
type & $c_0$ & $c_1$ & $c_2$  & $\sqrt{\chi^2/d.o.f}$ \\
\hline 
quadratic   & 1.866 & 0.67(4)   &      -      & 1.56 \\
quartic     & 1.866 & 0.40(9)   &  6.64(2.17) & 0.96 \\
Pad\'e      & 1.866 & -6.59(90) & -6.84(88)   & 0.76 \\
\thline
\end{tabular}
\end{center}
\end{table}

\begin{figure*}[htbp]
\begin{center}
\includegraphics[width=0.45\linewidth]{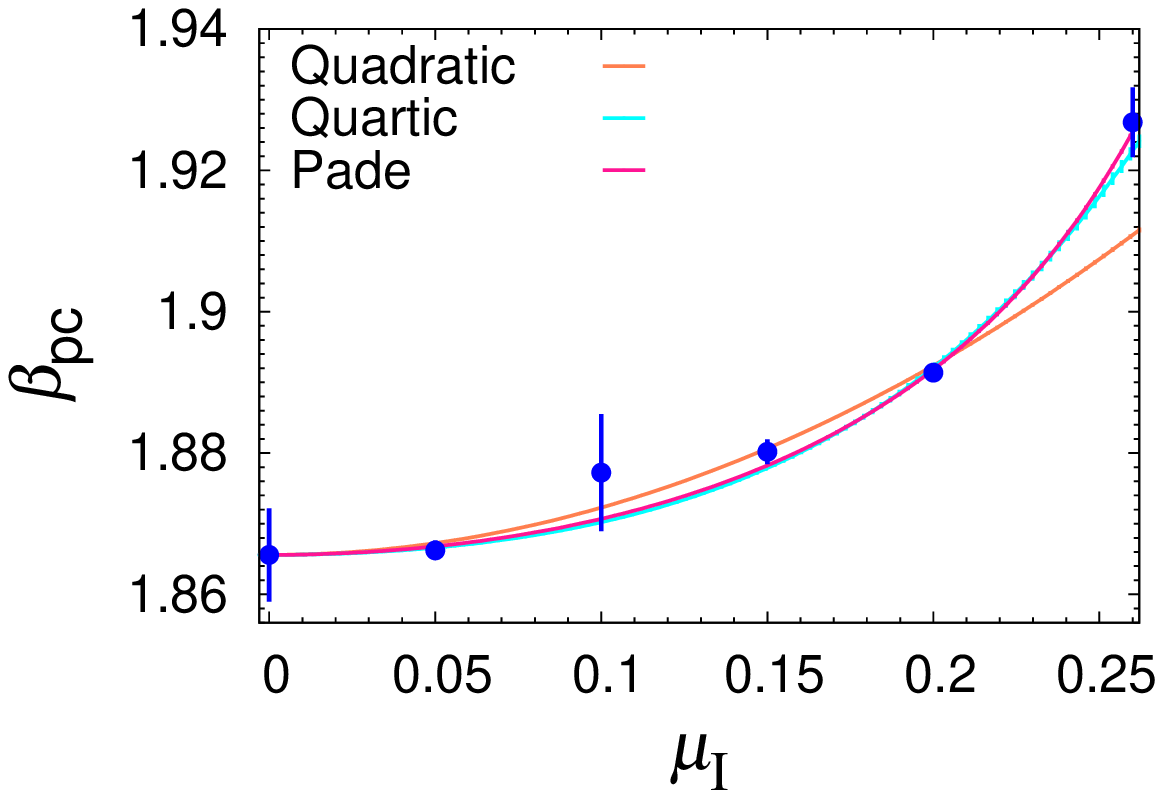}
\includegraphics[width=0.45\linewidth]{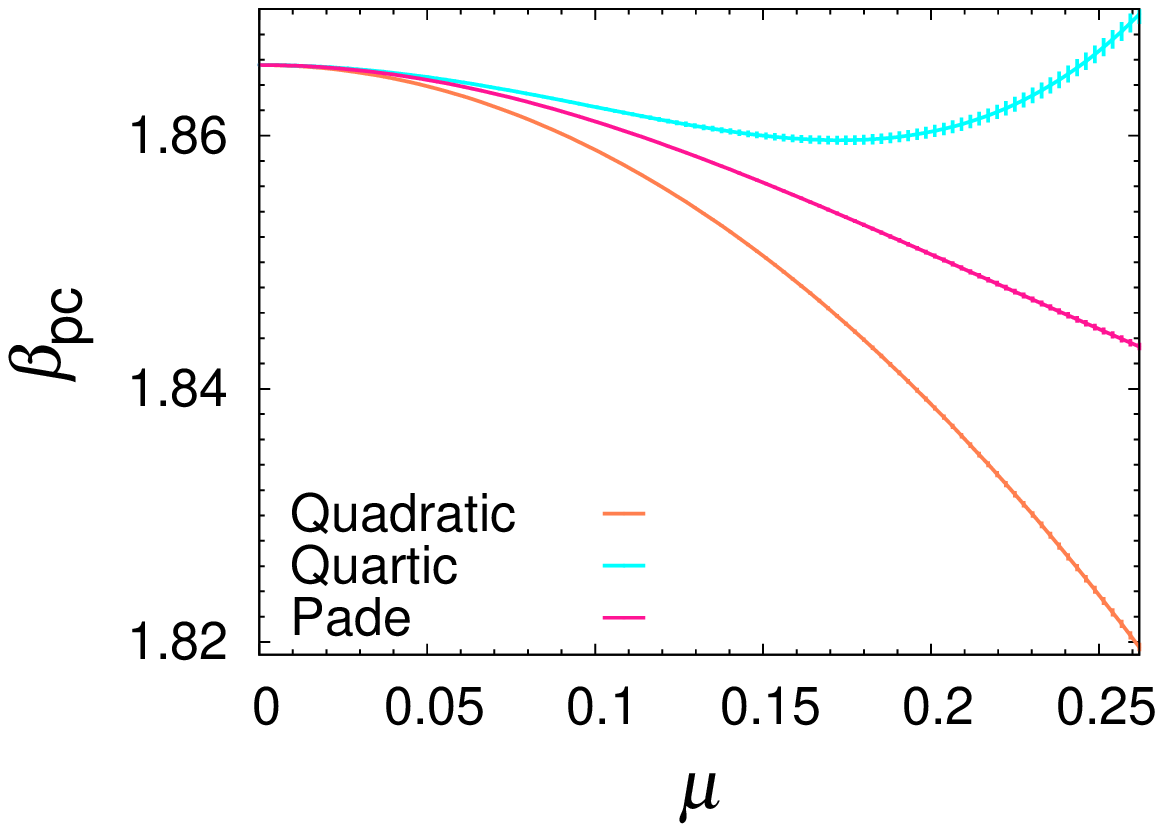}
\begin{minipage}{0.75\linewidth}
\caption{The pseudo-critical line $\beta_{pc}$ in the imaginary(left panel) 
and real(right panel) region.  
}\label{Feb2711fig1}
\end{minipage}
\end{center}
\end{figure*}

\begin{figure*}[htbp]
\begin{center}
\includegraphics[width=0.45\linewidth]{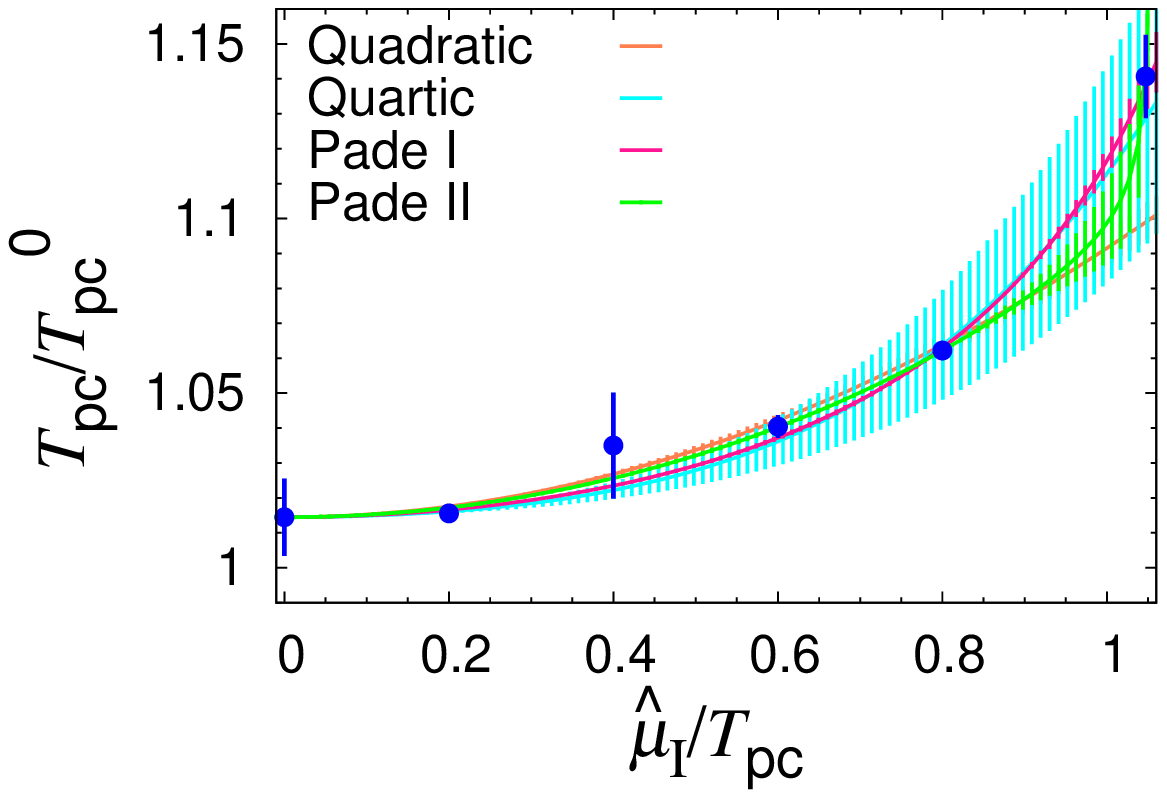}
\includegraphics[width=0.45\linewidth]{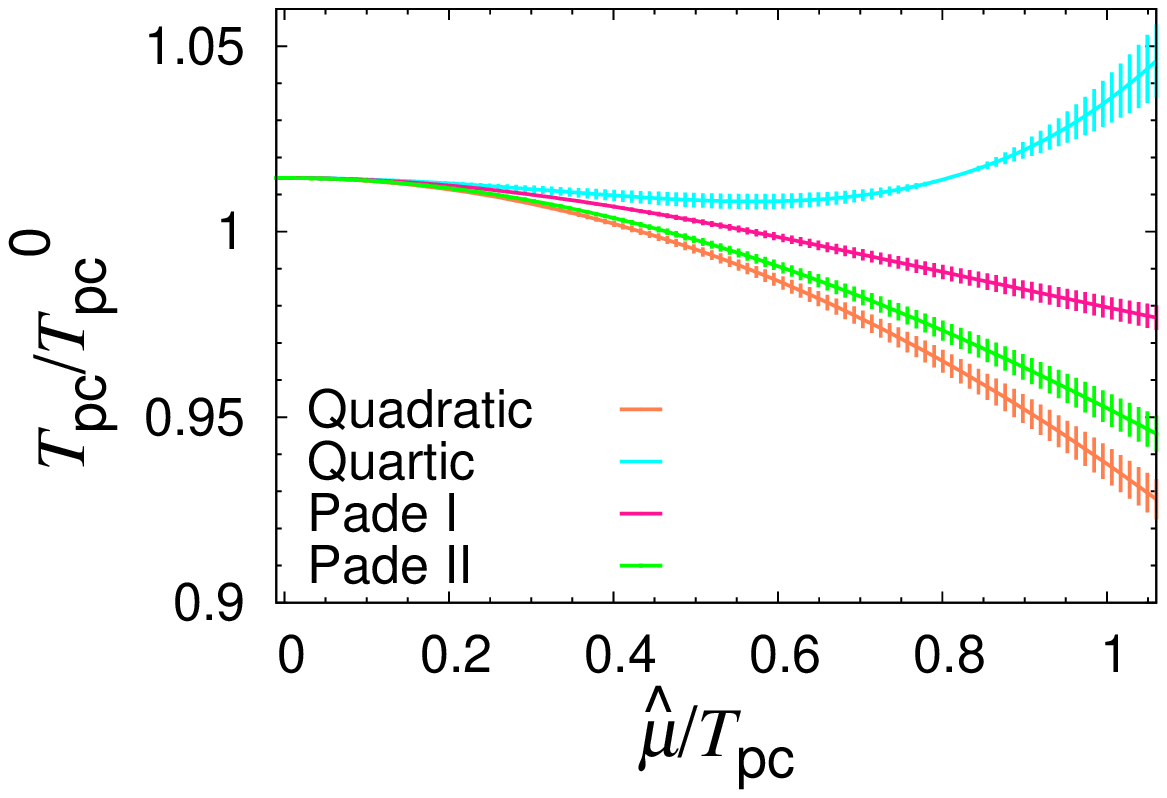}
\begin{minipage}{0.75\linewidth}
\caption{The pseudo-critical line in dimensionless physical unit in the imaginary 
(left panel) and real(right panel) region. 
}\label{Feb2711fig2}
\end{minipage}
\end{center}
\end{figure*}
The results are shown in Table~\ref{Feb2811tab1} and Fig.~\ref{Feb2711fig1}. 
The Pad\'e approximation and quartic function are better than the 
quadratic function because of non-$\mu_I^2$ contributions, as we have mentioned 
above. The Pad\'e approximation is slightly better than the quartic function in 
$\chi^2/d.o.f$, but the difference is quite small in the $\mu^2\le 0$ region. 
They are extended to the $\mu^2\ge 0$ plane through the analytic continuation 
$\mu_I^2 \to -\mu^2$. The result in the $\mu^2 \ge 0$ region depends on the fit 
functions in particular at large $\mu^2$. 

The pseudo-critical line $\beta_{pc}(\mu_I)$ can be transformed into the one 
in physical unit through the relation 
\begin{align}
T=\frac{1}{a(\beta) N_t}.
\label{Mar06eq1}
\end{align}
However, we use the data for $\beta$ dependence of $T/T_{pc}(0)$ obtained in 
Ref.~\cite{Ejiri:2009hq} instead of determining lattice spacings. Note that 
the value of $\beta_{pc}(0)$ slightly disagrees with the one obtained in 
Ref.~\cite{Ejiri:2009hq}, which causes the deviation of $T_{pc}/T_{pc}(0)$ at $\mu=0$ 
from 1 about 1\%  

\begin{table}[htbp]
\begin{center}
\begin{minipage}{\linewidth}
\caption{The coefficients of the fit functions for the pseudo-critical line.
}\label{Mar0611tab1}
\vspace{0.5cm}
\end{minipage}
\begin{tabular}{cccccc}
\thline
type & $d_0$ & $d_1$ & $d_2$  & $d_3$ & $\sqrt{\chi^2/d.o.f}$ \\
\hline 
quadratic      & 1.01 &  0.077(5)  & -         &  -         & 1.68 \\
quartic        & 1.01 &  0.039(12) & 0.060(19) &  -         & 1.00 \\
Pad\'e (I)     & 1.01 & -0.44(5)   & -0.49(5)  &  -         & 0.70 \\
Pad\'e (II)    & 1.01 & -0.89(10)  & -1.02(12) & 0.119(37)  & 0.55 \\
\thline
\end{tabular}
\end{center}
\end{table}

It should be noted that $T$ is not a linear function of $\beta$, 
therefore there is no need that the above functions can be used here. 
We consider quadratic, quartic functions and two types of the 
Pad\'e approximation. 
\begin{align}
\frac{T_{pc}}{T_{pc}^0} &= \sum_n d_n \left(\frac{\hat{\mu}_I}{T_{pc}}\right)^{2n}, 
\label{Mar0611eq3} \\
\frac{T_{pc}}{T_{pc}^0} &=  d_0 \frac{ 1 + d_1 (\hat{\mu}_I/T_{pc})^2}{1 + d_2 (\hat{\mu}_I/T_{pc})^2}\;\;\; \mbox{(Pad\'e (I))},
\label{Mar0611eq4}  \\
\left(\frac{T_{pc}}{T_{pc}^0}\right)^2 &=  d_0 \frac{ 1 + d_1 (\hat{\mu}_I/T_{pc})^2}{1 + d_2 (\hat{\mu}_I/T_{pc})^2 + d_3  (\hat{\mu}_I/T_{pc})^4 }
\label{May0711eq1} \nonumber \\ 
& \;\;\; \mbox{(Pad\'e (II))}, 
\end{align}
where $\mu_I = a \hat{\mu}_I$, and $\hat{\mu}_I$ is the imaginary chemical potential in 
physical unit. $T_{pc}^0$ and $T_{pc}$ are pseudo-critical temperatures at zero and 
finite chemical potentials.  As we have mentioned above, $d_0 (= T_{pc}/T_{pc}(0)$ at $\mu=0$) 
deviates from one with 1\% because of 
the disagreement of $\beta_{pc}(0)$ from Ref.~\cite{Ejiri:2009hq}. 
Here we added Pad\'e (II) defined in Eq.~(\ref{May0711eq1}), which  
was investigated  to fit the critical line for four flavors in Ref.~\cite{Cea:2010md}. 
Note that Eq.~(\ref{May0711eq1}) can be extended to $T_{pc}=0$ for the $\mu^2>0$ 
region~\cite{Cea:2010md}, although the definition range of the line is 
restricted by the RW endpoint. 
The results are shown in Table \ref{Mar0611tab1} and Fig.~\ref{Feb2711fig2}. 

The central values of the quadratic, quartic and Pad\'e (I) approximation show
similar behavior to $\beta_{pc}(\mu_I)$. The Pad\'e (II) are consistent with 
the quartic and Pad\'e (I) until $\mu_I/T_{pc}< 0.8$, and shows sharp rising 
for $0.8 < \mu_I/T_{pc}$. 
Considering the errors, the quadratic function still undershoots the obtained data, 
and the quartic function suffers from large errors. 
The two Pad\'e approximations reproduce the data well with small errors. The difference 
between the two Pad\'e approximations are observed for $0.8 < \mu_I/T$. 
Further investigations of this region would be important for better determination of the 
pseudo-critical line. 
 
Similar to $\beta_{pc}$, the deviation becomes larger with increasing $\hat{\mu}/T$ in 
the $\hat{\mu}^2\ge 0$ region and amounts to more than 10\% at $\hat{\mu}/T\sim 1$. 
The quartic function increases, and the other three functions decreases. 
The quartic function is completely different from the other three functions, 
although the quartic and Pad\'e (I) are almost same in the $\mu^2<0$ region. 

The quadratic function decreases the fastest, the Pad\'e (II) does the next, and 
the Pad\'e (I) overshoots the other two.
The curvature at $\hat{\mu}/T_{pc}=0$ of a power series of $(\hat{\mu}/\pi T_{pc})^2$ 
is often used to make comparisons with various studies~\cite{Philipsen:2008gf}. 
We obtain
\begin{align}
t_2 = \pi^2 d_1 = 0.38(12).
\end{align}
Here we employ the result for the quartic function. 
The result is smaller than the value obtained from staggered 
fermions~\cite{deForcrand:2002ci,Allton:2002zi} and standard Wilson 
fermions~\cite{Wu:2006su} and implies the pseudo-critical line decreases slower.

\section{Summary and outlook}

We have investigated the two-flavor QCD phase diagram in the lattice QCD. 
The imaginary chemical potential approach was employed in order 
to avoid the sign problem. The clover-improved Wilson action and renormalization-group 
improved gauge action was first applied to the imaginary chemical potential approach.
The simulation was performed on the $8^3\times 4$ lattice and at the intermediate 
quark mass. 
The simulation was performed for more than 150 points in the parameter 
$(\beta$, $\mu_I)$ plane. 
Considering the Polyakov loop, the imaginary chemical potential 
region of the phase diagram was examined. 

Obtained behaviors of the phase transitions are first order for the 
RW phase transition, second order for the RW endpoint and the 
crossover for the deconfinement. The corresponding order 
parameter for the phase transitions are the phase of the Polyakov 
loop for the RW phase transition and RW endpoint and the modulus 
of the Polyakov loop for the pseudo-critical line. 
The Polyakov loop modulus did not show a critical behavior on the RW phase 
transition line, while the Polyakov loop phase did not on the pseudo-critical line. 
We determined the pseudo-critical line from the susceptibility of the Polyakov 
loop modulus. We found a clear deviation from a linear dependence of the pseudo-critical 
line on $\mu_I^2$. 

The present calculation was performed with the intermediate quark mass and 
small lattice. The finite volume scaling analysis 
and quark mass-dependence analysis are necessary to confirm the present results. 
In particular, the order of the RW endpoint depends on the 
mass of the quark. The improvement on these points should be done in a future study. 

\section*{Acknowledgment}

We would like to appreciate for Yuji Sakai, Kouji Kashiwa, Hiroaki Kouno, and 
Masanobu Yahiro for discussions and valuable comments. 
KN thanks the XQCD-J collaboration, Shinji Motoki, Yoshiyuki Nakagawa, and 
Takuya Saito for discussions.

The simulation was performed on NEC SX-8R at RCNP, and NEC SX-9 at CMC, 
Osaka University, and HITACHI SR11000 and IBM Blue Gene/L at KEK.
This work was supported by Grants-in-Aid for Scientific Research 20340055 and 20105003.


\begin{thebibliography}{50}

\bibitem{Fodor:2010}
Z.~Fodor,
\newblock Proceedings of XL International Symposium on Multiparticle Dynamics,
  Univ. Antwerp, 2011  (2011), arXiv:.

\bibitem{DeTar:2011nm}
C.~DeTar,
\newblock (2011), arXiv:1101.0208.

\bibitem{Muroya:2003qs}
S.~Muroya, A.~Nakamura, C.~Nonaka, and T.~Takaishi,
\newblock Prog.Theor.Phys. {\bf 110}, 615 (2003), arXiv:hep-lat/0306031.

\bibitem{deForcrand:2010ys}
P.~de~Forcrand,
\newblock PoS {\bf LAT2009}, 010 (2009), arXiv:1005.0539.

\bibitem{Sakai:2008py}
Y.~Sakai, K.~Kashiwa, H.~Kouno, and M.~Yahiro,
\newblock Phys. Rev. {\bf D77}, 051901 (2008), arXiv:0801.0034.

\bibitem{Kouno:2009bm}
H.~Kouno, Y.~Sakai, K.~Kashiwa, and M.~Yahiro,
\newblock J. Phys. {\bf G36}, 115010 (2009), arXiv:0904.0925.

\bibitem{deForcrand:2002ci}
P.~de~Forcrand and O.~Philipsen,
\newblock Nucl. Phys. {\bf B642}, 290 (2002), arXiv:hep-lat/0205016.

\bibitem{D'Elia:2009qz}
M.~D'Elia and F.~Sanfilippo,
\newblock Phys. Rev. {\bf D80}, 111501 (2009), arXiv:0909.0254.

\bibitem{D'Elia:2009tm}
M.~D'Elia and F.~Sanfilippo,
\newblock Phys.Rev. {\bf D80}, 014502 (2009), arXiv:0904.1400.

\bibitem{deForcrand:2010he}
P.~de~Forcrand and O.~Philipsen,
\newblock Phys.Rev.Lett. {\bf 105}, 152001 (2010), arXiv:1004.3144.

\bibitem{D'Elia:2002gd}
M.~D'Elia and M.-P. Lombardo,
\newblock Phys.Rev. {\bf D67}, 014505 (2003), arXiv:hep-lat/0209146.

\bibitem{D'Elia:2004at}
M.~D'Elia and M.~P. Lombardo,
\newblock Phys.Rev. {\bf D70}, 074509 (2004), arXiv:hep-lat/0406012.

\bibitem{D'Elia:2007ke}
M.~D'Elia, F.~Di~Renzo, and M.~P. Lombardo,
\newblock Phys.Rev. {\bf D76}, 114509 (2007), arXiv:0705.3814.

\bibitem{Cea:2010md}
P.~Cea, L.~Cosmai, M.~D'Elia, and A.~Papa,
\newblock Phys.Rev. {\bf D81}, 094502 (2010), arXiv:1004.0184.

\bibitem{Cea:2009ba}
P.~Cea, L.~Cosmai, M.~D'Elia, C.~Manneschi, and A.~Papa,
\newblock Phys.Rev. {\bf D80}, 034501 (2009), arXiv:0905.1292.

\bibitem{Cea:2007vt}
P.~Cea, L.~Cosmai, M.~D'Elia, and A.~Papa,
\newblock Phys.Rev. {\bf D77}, 051501 (2008), arXiv:0712.3755.

\bibitem{Wu:2006su}
L.-K. Wu, X.-Q. Luo, and H.-S. Chen,
\newblock Phys. Rev. {\bf D76}, 034505 (2007), arXiv:hep-lat/0611035.

\bibitem{AliKhan:2000iz}
CP-PACS Collaboration, A.~Ali~Khan {\em et~al.},
\newblock Phys.Rev. {\bf D63}, 034502 (2001), arXiv:hep-lat/0008011.

\bibitem{Bonati:2009yg}
C.~Bonati, G.~Cossu, M.~D'Elia, A.~Di~Giacomo, and C.~Pica,
\newblock PoS {\bf LATTICE2008}, 204 (2008), arXiv:0901.3231.

\bibitem{Bernard:1993en}
C.~W. Bernard {\em et~al.},
\newblock Phys.Rev. {\bf D49}, 3574 (1994), arXiv:hep-lat/9310023.

\bibitem{Iwasaki:1996ya}
Y.~Iwasaki, K.~Kanaya, S.~Kaya, and T.~Yoshie,
\newblock Phys.Rev.Lett. {\bf 78}, 179 (1997), arXiv:hep-lat/9609022.

\bibitem{Roberge:1986mm}
A.~Roberge and N.~Weiss,
\newblock Nucl. Phys. {\bf B275}, 734 (1986).

\bibitem{Yang:1952be}
C.-N. Yang and T.~Lee,
\newblock Phys.Rev. {\bf 87}, 404 (1952).

\bibitem{Lee:1952ig}
T.~Lee and C.-N. Yang,
\newblock Phys.Rev. {\bf 87}, 410 (1952).

\bibitem{Iwasaki:1985we}
Y.~Iwasaki,
\newblock Nucl. Phys. {\bf B258}, 141 (1985).

\bibitem{Ejiri:2009hq}
WHOT-QCD Collaboration, S.~Ejiri {\em et~al.},
\newblock Phys.Rev. {\bf D82}, 014508 (2010), arXiv:0909.2121.

\bibitem{Sheikholeslami:1985ij}
B.~Sheikholeslami and R.~Wohlert,
\newblock Nucl. Phys. {\bf B259}, 572 (1985).

\bibitem{Philipsen:2008gf}
O.~Philipsen,
\newblock Prog.Theor.Phys.Suppl. {\bf 174}, 206 (2008), arXiv:0808.0672.

\bibitem{Allton:2002zi}
C.~Allton {\em et~al.},
\newblock Phys.Rev. {\bf D66}, 074507 (2002), arXiv:hep-lat/0204010.

\end{thebibliography}

\end{document}